\documentclass[preprint,preview,12pt]{elsarticle}

\usepackage{graphics}
\usepackage{graphicx}
\usepackage{amssymb}
\usepackage{latexsym}
\usepackage{makeidx}
\usepackage{multirow}
\usepackage{multicol}
\usepackage{float}
\usepackage{subfigure}
\usepackage{color, colortbl}
\usepackage{xcolor}

\usepackage{rotating}
\usepackage{caption}

\biboptions{square,numbers, comma, sort&compress}

\begin{document}

\begin{frontmatter}



\title{Report: Evolution of the PM2.5 concentration at \textit{Escuelas Aguirre} after \textbf{Madrid Central} implementation.}


\author[mcm]{Miguel C\'ardenas-Montes}
%\address[mcm]{CIEMAT, Department of Fundamental Research. \\Avda. Complutense 40. 28040. Madrid, Spain.}
\ead{miguel.cardenas.montes@gmail.com}

\begin{abstract}
In this report, the evolution of PM2.5 concentration at \textit{Escuelas Aguirre} monitoring station during the quarters of 2019 and the significance of the differences with the concentration of the quarters of previous years is analysed. 
These periods include the activation of \textbf{Madrid Central}, which is a major initiative for reducing motor traffic and the associated pollution at the city centre.
These periods correspond to the initialisation of, 2019Q1; and to tree fully-operational quarters of \textbf{Madrid Central}, 2019Q2, 2019Q3 and 2019Q4. 
\textit{Escuelas Aguirre} monitoring station is not inside \textbf{Madrid Central} but 2 km away.
The analysis of the PM2.5 concentration at this station allows discerning the rise of negative effects due to an increment of the motor traffic around the traffic restricted area. 
The analysis is based on the Binomial Sign Test for a Single Sample. 
The sample of signs is generated by subtracting the daily mean concentration of PM2.5 for the measurements of the period of two years: 20xx and 2019. 
Samples for the pair-wise comparisons for the year 2019 versus the years of the periods from 2010 to 2018 are evaluated.
An excess of positive or negative sign indicates that the differences between the measurements of concentrations in the comparison of two years are significant. 
Conversely, a balanced number of positive and negative signs points to no impact of the action. 
Since wind is an efficient pollution remover, the Binomial Sign Test is also applied to its intensity in the two-periods comparison. This allows discerning if the pollution reduction could be associated to a larger number of windy days in the 2019 quarter under comparison.
\end{abstract}

%\begin{keyword}
%GPU Computing \sep Performance \sep Correlation Function \sep Cosmology

%\end{keyword}
\end{frontmatter}

%\linenumbers

\section{Introduction}
Air pollution is one of the most critical health issue in urban areas. The scientific literature shows its relation with the population health \cite{Nel804,Linares2006,Diaz1999,AlberdiOdriozola1998}. For this reason, cities are implementing motor traffic restriction to the most pollutant vehicles in the centre of their central areas. Madrid city council has implemented \textbf{Madrid Central} by enlarging a previous area restricted to motor traffic at the centre of Madrid. Although this restriction was active at the beginning of 2019, only at begigning of 2019Q2 was fully operational. 

In the current report the impact on the PM2.5 concentration at \textit{Escuelas Aguirre} monitoring station of \textbf{Madrid Central} implementation is analysed. This station is outside of \textbf{Madrid Central}, 2 km away from the single monitoring station inside \textbf{Madrid Central}, \textit{Plaza del Carmen}. Unfortunately \textit{Plaza del Carmen} does not measure PM2.5 concentration.

The analysis of PM2.5 concentration at \textit{Escuelas Aguirre} monitoring station will allow discerning negative effect due an potential increment of motor traffic around \textbf{Madrid Central} area. The periods evaluated include the three first quarters of the years from 2010 to 2019. Thus the three quarters where \textbf{Madrid Central} is fully operative, 2019Q2, 2019Q3 and 2019Q4, are included in the study.

Samples for the pair-wise comparisons for the year 2019Q1, 2019Q2, 2019Q3 and 2019Q4 versus the years of the periods from 2010 to 2018 are evaluated (Eq. \ref{eq:pairwisecomp}). An excess of positive or negative sign indicates that the differences between the measurements of concentrations in the comparison of two years are significant; whereas a balanced distribution of signs points to no impact. The impact is evaluated through the Binomial Sign Test for a Single Sample.

\begin{eqnarray}
& [2010-01-01]-[2019-01-01] >0 := + \nonumber\\ 
& [2010-01-02]-[2019-01-02] <0 := - \nonumber\\ 
& [2010-01-03]-[2019-01-03] <0 := - \nonumber\\ 
& [2010-01-04]-[2019-01-04] >0 := + \nonumber\\ 
& \vdots \nonumber\\ 
& [2010-31-03]-[2019-31-03] <0 := - 
\label{eq:pairwisecomp}
\end{eqnarray}

%The rest of the document is organized as follows: Section \ref{section:relatedwork} summarizes related work and previous efforts made in this area.  
%An overview of the estimators of the Large-Scale Structure of the Universe and their computational complexity is presented in Sections \ref{section:MM:2PACF}, 
%The commonalities of the histogram construction on GPU are described in Section \ref{section:MM:commonalities}. 
%The weaknesses of floating-point representation are described in Section \ref{section:MM:weaknesses}. 
%The Bin-Recycling Strategy is briefly explained in Section \ref{section:MM:binrecycling}.
%The Results and the Analysis are shown in Section \ref{section:results}. Finally, the Conclusions are presented in Section \ref{section:conclusions}. 

%%%%%%%%%%%
\section{Methods and Materials\label{section:MM}}

\subsection{The Binomial Sign Test for a Single Sample\label{section:BinomialSignTest}}
The Binomial Sign Test for a Single Sample is based on the binomial distribution (Eq. \ref{eq:BinomialDistribution})\cite{Sheskin2004}. It assumes that any observation can be classified in one of the two mutually exclusive categories with probabilities $\pi_1$ and $\pi_2$. The evaluated hypothesis is if the proportion of the two categories are equal to a specific value, usually if they are equal. 

%%Hypothesis evaluated with test In an underlying population comprised of two categories that is represented by a sample, is the proportion of observations in one of the two categories equal to a specific value?

%%The basic assumption underlying the binomial distribution is that each of n independent observations (i.e., the outcome for any given observation is not influenced by the outcome for any other observation) is randomly selected from a population, and that each observation can be classified in one of k = 2 mutually exclusive categories.

%% page 197 and page 

\begin{equation}
P(x)= \frac{n!}{x!(n-x)!} \, \pi^x_1 \, \pi^{n-x}_2
\label{eq:BinomialDistribution}
\end{equation}
where $\pi_1 + \pi_2=1$, and therefore, $\pi_2=1-\pi_1$; $n$ is the sample size or the number of observations, and $x$ is the number of positive signs ($n-x$ is the number of negative signs).

The Null Hypothesis assumes that the true proportion of observations in any of the two categories are equal to $0.5$, $H_0: \, \pi_1=0.5$. Conversely, the (two-tailed) Alternative Hypothesis assumes that this is not true: \linebreak ${H_0: \, \pi_1 \ne 0.5}$.

In the current work the two categories correspond to the sign of the differences when subtracting the daily mean concentration of $NO_2$ for two years: 20xx-2019. Each sample includes the total daily mean values for Q1, 90 days; or 91 days for Q2. The test is repeated for each possible pair of years from 2010 to 2018 versus 2019. Thus, if daily mean values of Q1 for 2019 are significant lower than the year of comparison, then an excess of positive differences are obtained, $\pi_1>>0.5$. Conversely if the daily mean values of 2019 are significant larger than the year of comparison, then the excess corresponds to negative differences, $\pi_1<<0.5$. Finally, if the two years involved in the comparison behave similarly, then the probability of positive signs and negative signs wil be similar and close to $0.5$.

The computation of the confidence interval for $\pi_1$ is executed with the equation \ref{eq:BinomialDistributio:CI}.

\begin{equation}
p_1 - z_{\alpha/2} \sqrt{\frac{p_1 p_2}{n}} \le \pi_1 \le p_1 + z_{\alpha/2} \sqrt{\frac{p_1 p_2}{n}} 
\label{eq:BinomialDistributio:CI}
\end{equation}
where $p_1$ and $p_2$ are the proportion of positive and negative differences. For a confidence interval of 95\%, it must be computed with equation \ref{eq:BinomialDistributio:CI95}.

\begin{equation}
p_1 - 1.96 \cdot \sqrt{\frac{p_1 p_2}{n}} \le \pi_1 \le p_1 + 1.96 \cdot \sqrt{\frac{p_1 p_2}{n}} 
\label{eq:BinomialDistributio:CI95}
\end{equation}

When the interval does include the value $0.5$, it will mean that the differences are not significant for a confidence level of 95\% (p-value under 0.05). In this case, the Null Hypothesis can not be rejected, and therefore, no impact can be attributed to the actuation. 

The Binomial Sign Test for a Single Sample has been implemented in R with package MASS \cite{VenablesRipley}. It allows compute the confidence interval for $\pi_1$.

This methodology is applied not only to the PM2.5 concentration measured at \textit{Escuelas Aguirre} monitoring station, but also to the wind velocity measured at Madrid Airport by AEMET. Wind is an efficient pollution remover and for this reason its presence is also evaluated. This allows discerning if the reduction of the pollution is due to a reduction of the emission or due to the presence of more windy days in the period analysed. 

%%%%%%%%% 
\section{Results and Discussion\label{section:results}}

\begin{figure*}
{\renewcommand{\arraystretch}{1.0}
\rotatebox{90}{
\begin{minipage}[c][][c]{\textheight}
\centering
  \subfigure[2019Q1]{
    \includegraphics[width=0.47\textwidth]{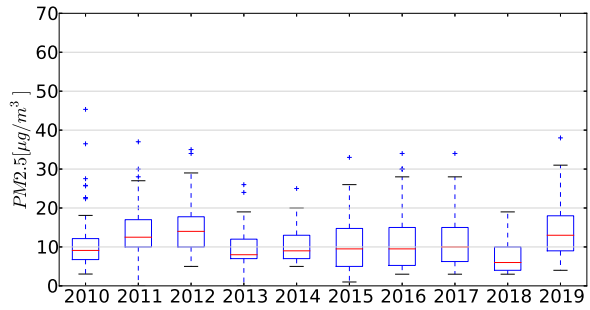}
  \label{figure:boxplot:2019Q1:PM25:EA}}
  \subfigure[2019Q2]{
    \includegraphics[width=0.47\textwidth]{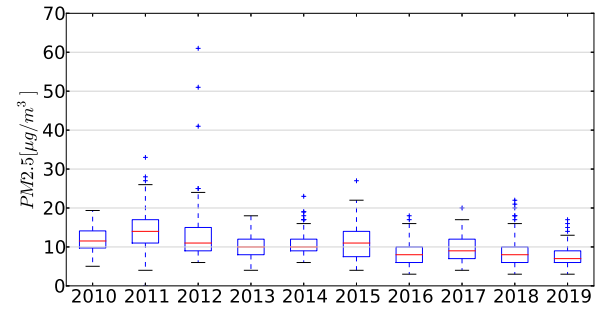}
  \label{figure:boxplot:2019Q2:PM25:EA}}
  \subfigure[2019Q3]{
    \includegraphics[width=0.47\textwidth]{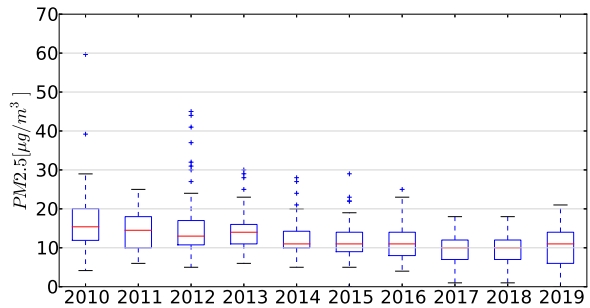}
  \label{figure:boxplot:2019Q3:PM25:EA}}
  \subfigure[2019Q4]{
    \includegraphics[width=0.47\textwidth]{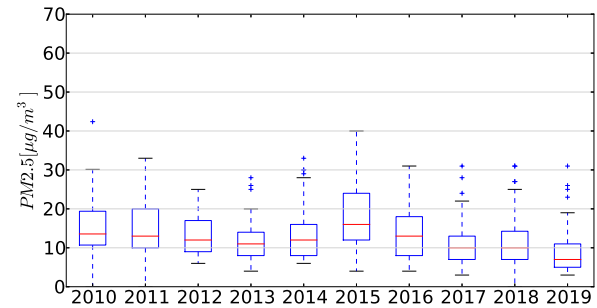}
  \label{figure:boxplot:2019Q4:PM25:EA}}
  \caption{Boxplot with the mean daily concentration of PM2.5 for the periods 2019Q1 (Fig. \ref{figure:boxplot:2019Q1:PM25:EA}), 2019Q2 (Fig. \ref{figure:boxplot:2019Q2:PM25:EA}), 2019Q3 (Fig. \ref{figure:boxplot:2019Q3:PM25:EA}), and 2019Q4 (Fig. \ref{figure:boxplot:2019Q4:PM25:EA}) at \textit{Escuelas Aguirre} monitoring station. Red horizontal line shows the median of the values for the period 2010-2019.} 
\label{figure:boxplot:2019:PM25:EA}
\end{minipage}
        }
}
\end{figure*}

In Fig. \ref{figure:boxplot:2019:PM25:EA}, the mean daily concentration of PM2.5 for the periods 2019Q1, 2019Q2, 2019Q3 and 2019Q4 at \textit{Escuelas Aguirre} are shown. As can be observed, measurements for the quarters of 2019 are similar to previous years, with slight higher values for the observartions of period 2019Q1 (Fig. \ref{figure:boxplot:2019Q1:PM25:EA}), and a smooth tend to lower values for the periods 2019Q2 (Fig. \ref{figure:boxplot:2019Q2:PM25:EA}), 2019Q3 (Fig. \ref{figure:boxplot:2019Q3:PM25:EA}), and 2019Q4 (Fig. \ref{figure:boxplot:2019Q4:PM25:EA}). 

\begin{figure*}
%{\renewcommand{\arraystretch}{1.0}
%\rotatebox{90}{
%\begin{minipage}[c][][c]{\textheight}
\centering
  \subfigure[2019Q1]{
    \includegraphics[width=0.39\textwidth]{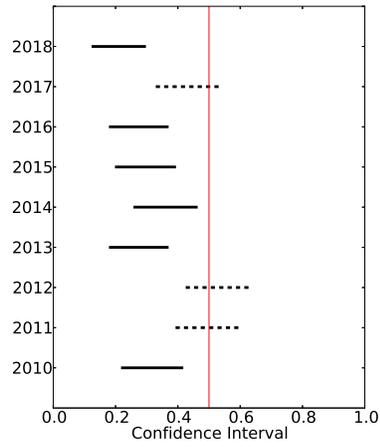}
  \label{figure:ci:binomial:EA:2019Q1}}
  \subfigure[2019Q2]{
    \includegraphics[width=0.39\textwidth]{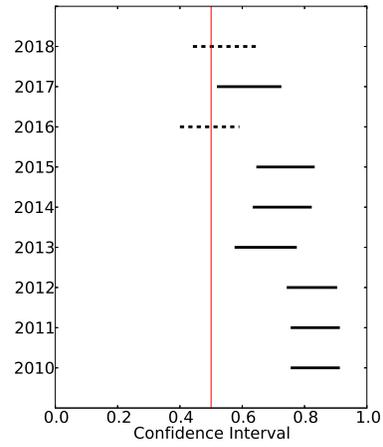}
  \label{figure:ci:binomial:EA:2019Q2}}\\
  \subfigure[2019Q3]{
    \includegraphics[width=0.39\textwidth]{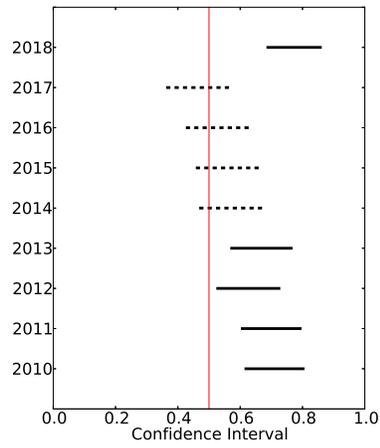}
  \label{figure:ci:binomial:EA:2019Q3}}
  \subfigure[2019Q4]{
    \includegraphics[width=0.39\textwidth]{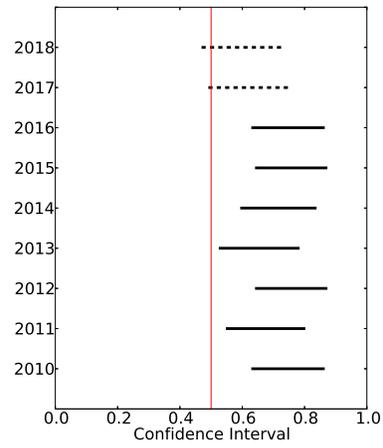}
  \label{figure:ci:binomial:EA:2019Q4}}
\caption{Confidence intervals for the pair-wise comparisons of PM2.5 concentration with binomial sign test for a single sample for \textit{Escuelas Aguirre} monitoring station.}
\label{figure:ci:binomial:EA}
%\end{minipage}
%        }
%}
\end{figure*}

In Fig. \ref{figure:ci:binomial:EA:2019Q1} the confidence interval of $\pi_1$ for the pair-wise comparisons of 2019Q1 are shown. As can be appreciated, in most of the cases the differences are significant for a confidence level of 95\% (p-value under 0.05), being the observations of 2019Q1 significantly worse than the observations of the years: 2010, 2013, 2014, 2015, 2016, and 2018; therefore, the concentration values are significantly higher in 2019Q1 than in the corresponding comparison. This means that the differences are unlikely to have occurred by chance with a probability of 95\%. Conversely, the differences with the observations of years: 2011, 2012 and 2017 are not significant. 

For the period 2019Q2 (Fig. \ref{figure:ci:binomial:EA:2019Q2}), the confidence intervals of $\pi_1$ indicate that most the differences are significant for a confidence level of 95\% (p-value under 0.05), being the observations of 2019Q2 lower than the observations of the years in the period from 2010 to 2015, and 2017. Conversely, the differences are not significant for a confidence level of 95\% for the pair-wise comparisons with the years: 2016 and 2018.

Regarding the period 2019Q3 (Fig. \ref{figure:ci:binomial:EA:2019Q3}), four pair-wise comparisons are not significant for a confidence level of 95\% (p-value under 0.05), those corresponding to years from 2014 to 2017; whereas for the years from 2010 to 2013, and the year 2018 the differences are significant for a confidence level of 95\% (p-value under 0.05), which means that the differences are unlikely to have occurred by chance with a probability of 95\%. The concentration of PM2.5 for this period is significant lower than the mentioned years.

Finally for the period 2019Q4 (Fig. \ref{figure:ci:binomial:EA:2019Q4}), seven pair-wise comparisons are significant for a confidence level of 95\% (p-value under 0.05), therefore, concentration values for 2019Q4 are lower than the concentration values of years in the periods from 2010 to 2016. Only the pair-wise comparisons corresponding to the comparisons with years 2017 and 2018 are not significant for a confidence level of 95\%. 

A major remover of air pollution in large urban areas is the wind. Therefore, it must be ascertained if the reduction of PM2.5 observed in pair-wise comparisons for periods 2019Q2, 2019Q3 and 2019Q4 are due to a reduction of emissions or due to more windy days in these periods in comparison with the previous years. For this purpose, the binomial sign test for a single sample can also be used, but analysing the wind velocity observations. 

The pair-wise comparisons are undertaken following the schema of Eq. \ref{eq:pairwisecomp} by using the wind velocity provided by AEMET. More windy days in the periods 2019Q2, 2019Q3 and 2019Q4 lead to quarantine the reduction of PM2.5 concentration in \textit{Escuelas Aguirre} monitoring station due to a positive frontier effect of \textbf{Madrid Central}. Oppositely, when the differences in the wind velocity observations do not show significant differences or the significant differences are in favour of less windy days in the periods 2019Q2, 2019Q3 and 2019Q4, a positive reduction of PM2.5 concentration in \textit{Escuelas Aguirre} monitoring station can be claimed.

\begin{figure*}
%{\renewcommand{\arraystretch}{1.0}
%\rotatebox{90}{
%\begin{minipage}[c][][c]{\textheight}
\centering
  \subfigure[2019Q1]{
    \includegraphics[width=0.39\textwidth]{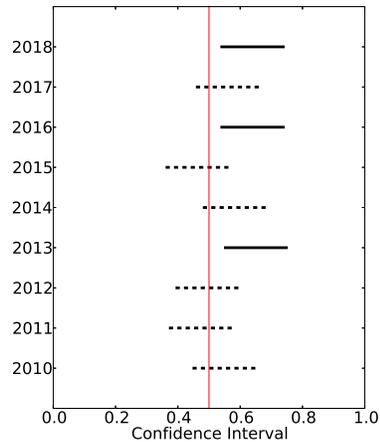}
  \label{figure:ci:binomial:wind:2019Q1}}
  \subfigure[2019Q2]{
    \includegraphics[width=0.39\textwidth]{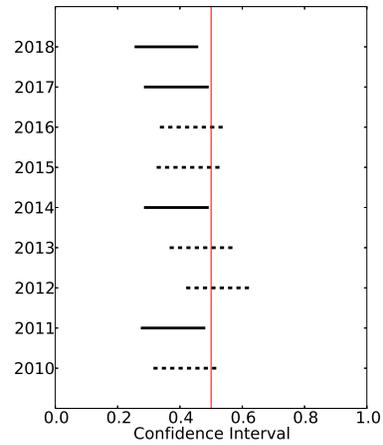}
  \label{figure:ci:binomial:wind:2019Q2}}
  \subfigure[2019Q3]{
    \includegraphics[width=0.39\textwidth]{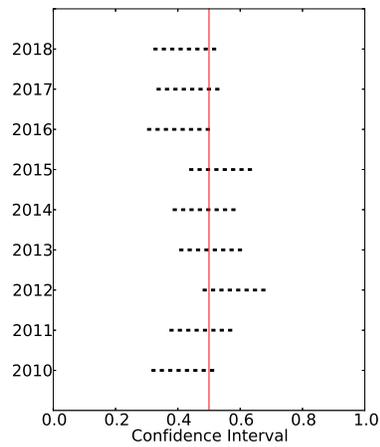}
  \label{figure:ci:binomial:wind:2019Q3}}
  \subfigure[2019Q4]{
    \includegraphics[width=0.39\textwidth]{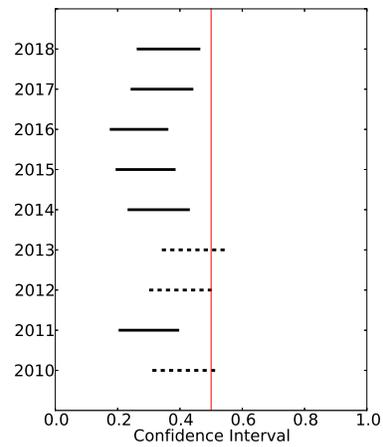}
  \label{figure:ci:binomial:wind:2019Q4}}
\caption{Confidence intervals for the pair-wise comparisons wind velocity  at Madrid Airport with binomial sign test.}
\label{figure:ci:binomial:wind}
%\end{minipage}
%        }
%}
\end{figure*}

In Figs. \ref{figure:ci:binomial:wind:2019Q2}, \ref{figure:ci:binomial:wind:2019Q3} and \ref{figure:ci:binomial:wind:2019Q4}, the confidence intervals of $\pi_1$ for the pair-wise comparisons of the wind velocity observations are shown. For the period 2019Q2 (Fig. \ref{figure:ci:binomial:wind:2019Q2}), no significant differences are shown for the pair-wise comparisons with the years: 2010, 2012, 2013, 2015 and 2016; and as a consequence, a positive reduction of PM2.5 concentration not associated to more windy days for the years 2010, 2012, 2013, and 2015 can be claimed (Fig. \ref{figure:ci:binomial:EA:2019Q2}). 

The previous analysis for the period 2019Q3 indicates that differences in the pair-wise comparisons for windy days are not significant for any comparison (Fig. \ref{figure:ci:binomial:wind:2019Q3}). Therefore, all the positive reductions of PM2.5 concentration in \textit{Escuelas Aguirre} monitoring station can be claimed (Fig. \ref{figure:ci:binomial:EA:2019Q3}). %For the periods 2019Q2 and 2019Q3, 9 positive pair-wise comparisons can be claimed from a total of 18 cases. 

Except for the years: 2010, 2012 and 2013, in the fourth quarter of 2019, the days have been significantly more windy in most of the years comparisons (Fig. \ref{figure:ci:binomial:wind:2019Q4}). Consequently from the positive reductions in the PM2.5 concentration observed in most of the comparisons with 201Q4 period (Fig. \ref{figure:ci:binomial:EA:2019Q4}) only those observed for the years 2010, 2012 and 2013 can be claimed as not linked to more windy days in 2019Q4. 

For the periods 2019Q2, 2019Q3 and 2019Q4, 12 positive pair-wise comparisons ---in favour of a reduction of PM2.5 not related with more windy days--- can be claimed from a total of 27 comparisons. 

The previous analysis can be also applied to the period 2019Q1, when \textbf{Madrid Central} was not fully operational. The pair-wise comparisons for the windy velocity observations indicate that the differences are not significant for the years: 2010, 2011, 2012, 2014, 2015 and 2017 (Fig. \ref{figure:ci:binomial:wind:2019Q1}). 

The comparisons between 2019 and the years 2010, 2014 and 2015, indicate that the PM2.5 concentration is significant worse ---in 2019--- than those years, without worse meteorological conditions (Fig. \ref{figure:ci:binomial:EA:2019Q1}). As can be appreciated, a change in the tendency is observed from 2019Q1 to 2019Q2 in coincidence with the \textbf{Madrid Central} activation. 

%For the years: 2010, 2014 and 2015, the PM2.5 concentration is significant worse than the previous years without worse meteorological conditions (Fig. \ref{figure:ci:binomial:EA:2019Q1}). As can be appreciated, a change in the tendency is observed from 2019Q1 to 2019Q2 in coincidence with the \textbf{Madrid Central} activation. 

%Furthermore the pair-wise comparisons for PM2.5 concentration is in most of the case significant worse that their equivalent periods for previous years.

%%%%%%%%%%
\section{Conclusions}\label{section:conclusions}
In this report the evolution of the PM2.5 concentration in \textit{Escuelas Aguirre} monitoring station after the activation of \textbf{Madrid Central} ---a restricted motor-traffic area--- is analysed. The PM2.5 concentration for the quarters of 2019 in comparison with their equivalent periods of the years from 2010 to 2018 is included in the analysis. For this purpose, the Binomial Sign Test for a Single Sample has been used. \textit{Escuelas Aguirre} monitoring station is outside the restricted motor-traffic area, at 2 km of the border. The contribution of the windy days as an efficient pollution remover has been also considered in the study. From the analysis it can rejected any increment of the PM2.5 concentration due the activation of \textbf{Madrid Central}, even if the positive effect of windy days is considered.

\end{document}